\documentclass[preprint,journal]{vgtc}       

\ifpdf
  \pdfoutput=1\relax                   
  \usepackage{graphicx}                
  \DeclareGraphicsExtensions{.pdf,.png,.jpg,.jpeg} 
\else
  \usepackage{graphicx}                
  \DeclareGraphicsExtensions{.eps}     
\fi%

\graphicspath{{figures/}{pictures/}{images/}{./}} 

\usepackage{microtype}                 
\PassOptionsToPackage{warn}{textcomp}  
\usepackage{textcomp}                  
\usepackage{mathptmx}                  
\usepackage{times}                     
\usepackage{cite}                      
\usepackage{tabu}                      
\usepackage{booktabs}                  

\preprinttext{Author preprint. To appear in IEEE Transactions on Visualization and Computer Graphics.}

\onlineid{0}

\vgtccategory{Research}
\vgtcpapertype{please specify}

\title{Understanding Data Visualization Design Practice}

\author{Paul Parsons}
\authorfooter{
\item
Paul Parsons is with Purdue University. E-mail: parsonsp@purdue.edu.
}

\shortauthortitle{Parsons: Understanding Data Visualization Design Practice}

\abstract{Professional roles for data visualization designers are growing in popularity, and interest in relationships between the academic research and professional practice communities is gaining traction. However, despite the potential for knowledge sharing between these communities, we have little understanding of the ways in which practitioners design in real-world, professional settings. Inquiry in numerous design disciplines indicates that practitioners approach complex situations in ways that are fundamentally different from those of researchers. In this work, I take a practice-led approach to understanding visualization design practice on its own terms. Twenty data visualization practitioners were interviewed and asked about their design process, including the steps they take, how they make decisions, and the methods they use. Findings suggest that practitioners do not follow highly systematic processes, but instead rely on situated forms of knowing and acting in which they draw from precedent and use methods and principles that are determined appropriate in the moment. These findings have implications for how visualization researchers understand and engage with practitioners, and how educators approach the training of future data visualization designers. %
} 


\vgtcinsertpkg

\begin{document}


\firstsection{Introduction}

\maketitle


Professional roles for data visualization designers are growing in popularity \cite{meeks_2019_2019,noauthor_data_2019}, and interest in relationships between the researcher and practitioner communities is gaining traction. However, despite the potential for knowledge production and sharing across these communities, we have little understanding of the ways in which practitioners design in real-world settings. The knowledge that has been generated about data visualization design has largely come from research settings rather that settings involving professional practice. Popular models of visualization design (e.g., \cite{munzner_nested_2009,sedlmair_design_2012,mckenna_design_2014}) have come from researchers reflecting on their own design work primarily in research settings. While scholarship on practitioner-oriented issues has been gaining traction in recent years \cite{bigelow_iterating_2017,bigelow_reflections_2014,hoffswell_techniques_2020,mendez_bottom-up_2017,walny_data_2020,parsons_what_2020,alspaugh_futzing_2019}, only a small number of studies have been grounded in everyday design practice (e.g., \cite{bigelow_reflections_2014,hoffswell_techniques_2020,mendez_bottom-up_2017,walny_data_2020,alspaugh_futzing_2019}). 

Multiple design-related fields have embraced a broad view of design that includes practice perspectives in addition to traditional research ones. This view is often accompanied by a recognition that there are barriers to knowledge production and use between the academic and practitioner communities. This is true across emergent design disciplines such as interaction design \cite{goodman_understanding_2011, Stolterman2008, kuutti_turn_2014}, user experience design \cite{kou_practice-led_2019}, and instructional design \cite{boling_core_2017}, and in more traditional disciplines such as architecture \cite{schon_reflective_1983}. In HCI scholarship, the idea of a ``gap'' existing between research and practice has been discussed for many years \cite{Stolterman2008,Rogers2004,kou_practice-led_2019,velt_translations_2020,colusso_translational_2017,beck_theory-practice_2018,goodman_understanding_2011,zimmerman_research_2007}. Although the visualization community has held events in recent years involving both researchers and practitioners (e.g., \cite{noauthor_workshop_2019,noauthor_workshop_2020,noauthor_visualization_2020,noauthor_information+_nodate}), the relationship itself has not been the focus of much scholarly inquiry. If there is a gap between visualization research and practice, it cannot be addressed adequately until the practices of professional visualization designers are well understood.

Based on findings in various design disciplines, it is reasonable to anticipate value in studying data visualization design practice as distinct from data visualization research. Doing so depends on recognizing professional practice as an activity with its own ways of knowing and acting that may be fundamentally different from those in academic settings. The work presented here is one attempt to take this approach. In this paper, I first draw from research in other design disciplines to characterize design practice, discuss its relationship to design research, and describe a brief history of design methods in a broad sense. Two studies are then presented. First, 87 data visualization practitioners were surveyed to understand their familiarity with popular design methods and the frequency with which they use them in their own work. Second, 20 professional practitioners were interviewed, in which they were asked about their design process, strategies, and use of methods and concepts that guide their practice. Issues relating to the use of precedent and experience in the design process are surfaced, including the reliance on situated planning and knowing, and the kinds of design knowledge that practitioners appreciate and use.

This work contributes a theoretical framing and language for studying professional data visualization practice, especially in relation to how practitioners confront the complexity of real-world design situations in ways that differ from how researchers do. This framing and language can be helpful for considering how knowledge is produced and shared across the research and practice communities. Based on the interview findings, this work opens up new questions about how practitioners design in real-world settings, how findings from research studies are used in their practice, and how researchers might structure their attempts to understand and influence design practice. 

\section{Defining Design Practice}
The concept of ``practice'' can take on different meanings, so it is important to define its use here. HCI scholars have attempted to bring more clarity to the concept of practice for many years. For instance, Kuutti and Bannon \cite{kuutti_turn_2014} argued a need for a ``turn to practice'' in HCI, contrasting a practice paradigm with what they called the dominant interaction paradigm. Within the interaction paradigm, the scope of analysis involves human actions being influenced by means of a technological intervention, largely abstracted from a particular time and place. Within the practice paradigm, issues relating to the material and cultural environment are considered, along with everything related to and interwoven in the performance of an activity \cite{kuutti_turn_2014}. Goodman et al. \cite{goodman_understanding_2011} draw from practice theories in sociology, viewing practice as comprising the activities, experiences, and contexts of individuals operating within and influenced by technical systems, organizational structures, tools, and knowledge. Schmidt \cite{rossitto_concept_2014} provides a detailed review of the concept of practice, including historical usage trends and some philosophical considerations. Schmidt notes that the point of employing the concept of practice is to focus on the unity of action in work, including the reciprocity of general knowledge and contingent action. Practices are neither fully ad hoc action nor fully abstracted procedures. Rather, practices are contingent activities that are performed with regularity. As Kuutti and Bannon \cite{kuutti_turn_2014} note, they are ``ways how things get done, continuously produced and reproduced.''

Drawing from this prior work, practice is defined in this paper as activity that is done with regularity, based on common rules and principles; involving professional skill and competence; and including experiences and contextual factors influential in professional settings, such as clients, budgets, timelines, and organizational structures. The focus in this paper includes identifiable aspects of design activity that are aimed at creating particular visualization artifacts in the context of professional practice as defined above. Unlike research settings, knowledge production is usually not a primary goal in practice settings. The goal is rather to create artifacts that satisfy the needs of a client within the constraints of the practice setting.


\section{Research-Practice Relationships}
In his seminal work, Donald Schön critiqued the academic view of the professions, calling for an ``inquiry into the epistemology of practice'' \cite{schon_reflective_1983}, including the study of what practitioners actually do rather than what researchers think they do. He drew a clear distinction between scholarly knowledge that is valued in academia and the ``practical competence and professional artistry'' that is important for practice. His critique noted that researchers tend not to value this kind of competence and artistry, largely because it cannot be abstracted and codified. This has to do with the nature of tacit knowledge \cite{polanyi_tacit_1966} that is prevalent in the professions, and Schön notes that practitioners usually ``know more than they can say'' \cite{schon_reflective_1983}. Much of the tacit knowledge of professionals is not always amenable to explicit articulation, often making it difficult for researchers to know how to study such phenomena \cite{polanyi_tacit_1966}.  

Among design-oriented fields, the ways in which academics view design practice are often not examined critically. Roedl and Stolterman \cite{roedl_design_2013} examined proceedings from the 2011 CHI conference and identified papers that aimed to support design practice. They found it was common to over-generalize design situations---meaning that authors talk about ``designers'' or the ``design process'' in vague and generic ways that do not distinguish between important contextual factors and practical challenges of professional design work. Regarding how researchers view practice, they found that ``conceptualizations were overly simplistic or biased towards the way design works in a research setting.'' Gray et al. \cite{gray_reprioritizing_2014} later critiqued conceptions of practice common in the HCI literature, arguing that the research community relies on a ``projected practice community'' that may not be representative of actual practice. 



Although HCI has a history of viewing research-practice relationships as objects of study (e.g., \cite{goodman_understanding_2011,Stolterman2008,roedl_design_2013,gray_reprioritizing_2014,wolf_dispelling_2006,colusso_translational_2017}), VIS has not seen a parallel degree of effort. There have been attempts in recent years, however, to bring together researchers and practitioners through various events. These include recurring workshops at VIS, such as VisinPractice \cite{noauthor_workshop_2019}, VisGuides \cite{noauthor_workshop_2020}, and VisComm \cite{noauthor_visualization_2020}, and external events such as the Information+ Conference \cite{noauthor_information+_nodate}. Other initiatives include the \textit{Data Stories} podcast \cite{bertini_data_nodate}, and the blog \textit{Multiple Views: Visualization Research Explained} \cite{noauthor_multiple_2019}. There is also the VisGuides website \cite{noauthor_visguidesorg_nodate} that provides a forum for researchers and practitioners to discuss visualization guidelines. There has recently been the creation of the Data Visualization Society \cite{noauthor_data_2019}, which has engaged a number of academic researchers on its advisory board. It appears that there is increasing interest in bringing researchers and practitioners together to learn from one another. This is an important step; however, in line with work in other disciplines, there is a need to go beyond creating these connections to actually studying them as a form of scholarship. While it is likely that some form of ``gap'' exists between these communities, it has not yet been characterized through scholarly inquiry.

\begin{figure}[h]
  \centering
  \includegraphics[width=0.9\columnwidth]{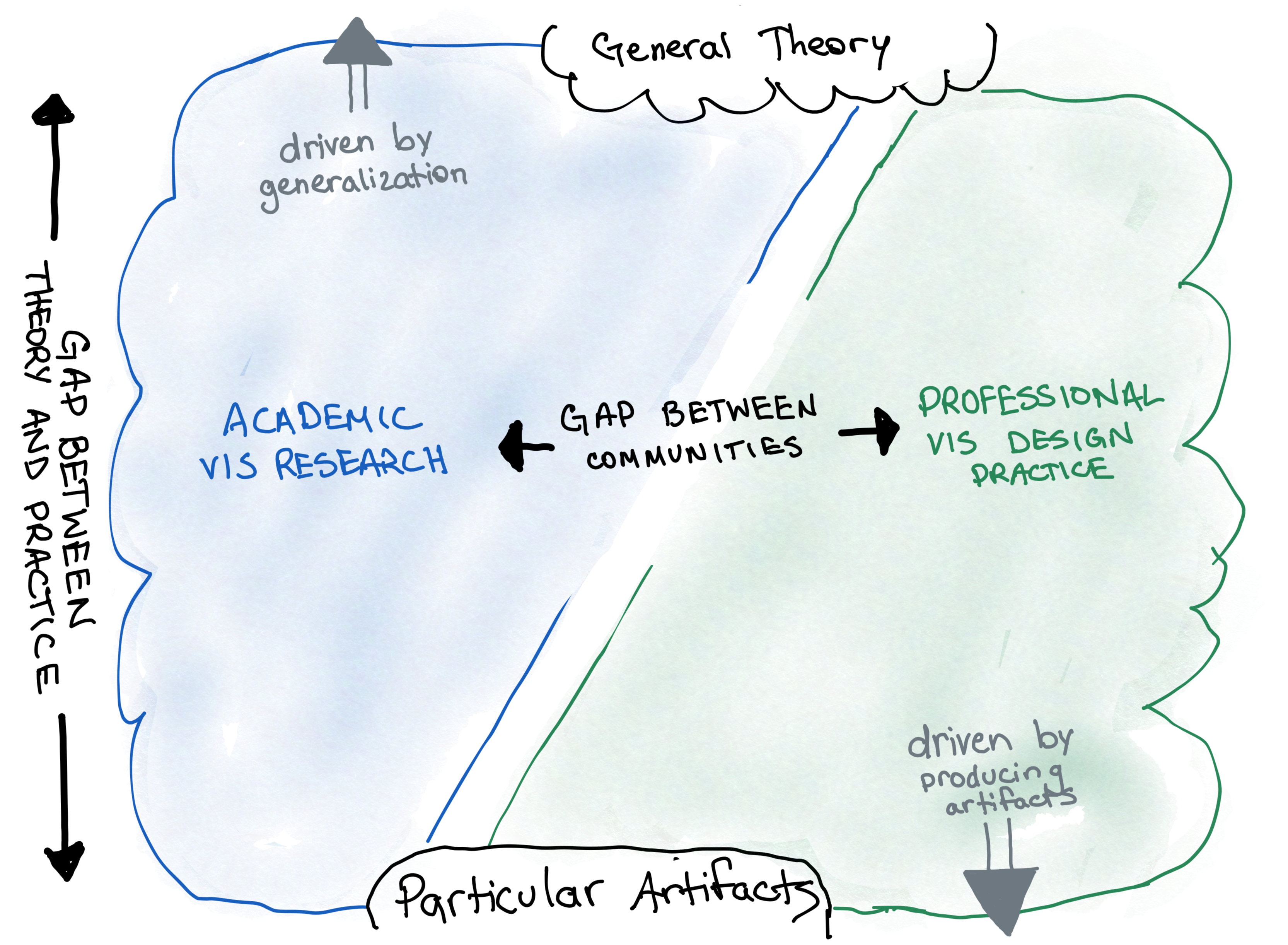}
  \caption{Dual gap model between research and professional practice communities, adapted from Velt et al. \cite{velt_translations_2020}.}~\label{fig:gap}
\end{figure}

Figure \ref{fig:gap} depicts a gap between the research and professional practice communities, adapted from Velt et al.'s \cite{velt_translations_2020} work in HCI and inspired by previous work on the nature of design knowledge \cite{nelson_design_2012,hook_strong_2012}. On the left is the academic research community, which is largely driven by producing general knowledge. On the right is the professional practice community, which is largely driven by producing particular artifacts. This depiction is somewhat of a simplification, as researchers are also sometimes concerned with creating specific artifacts, and practitioners are sometimes concerned with developing general knowledge---e.g., in the form of design patterns, guidelines, and other intermediate-level forms of knowledge \cite{hook_strong_2012}. Velt et al.'s \cite{velt_translations_2020} model proposes two gaps---one between two different communities and the other between theory and particular design instances. The second gap may exist within each and across both communities---for instance, there may be particular visualization techniques without any clear connection to theory, and there may be theory without clear influence on particular artifacts. 

\subsection{Practice vs. Application}
The VIS community has grappled with how application-oriented work fits in the research landscape. Researchers have engaged in discussions on how to effectively evaluate application papers \cite{weber_apply_2017}, and the literature has seen a growing number of ``systems'', ``application'', and ``design studies'' papers \cite{lee_broadening_2019,kerren_process_2008}. Application-oriented research and ``practice'' are sometimes discussed together in the visualization literature as if they are the same (e.g., \cite{hentschel_applied_2018}). This view relies on a key assumption about the relationship between research and practice---that the flow of knowledge is largely unidirectional, where researchers create it and practitioners employ it. This assumption does not acknowledge the distinct epistemology of practice. While scientific knowledge certainly plays a role in design, it is not sufficient for good design \cite{Stolterman2008}. Rather, designers rely on a host of personal and situated factors, along with more formal types of knowledge, to engage appropriately with the complexity of design practice. Buchanan \cite{buchanan_wicked_1992} articulates how widespread this assumption has been, noting that ``each of the sciences that have come into contact with design has tended to regard design as an `applied' version of its own knowledge'', emphasizing the mistake of viewing design as simply a ``practical demonstration'' of scientific findings. To accurately understand design practice, it must be viewed as distinct from application---an activity having its own forms of competence and processes for knowledge production and use.


\subsection{Role of Visualization Design Studies}
Visualization researchers have contributed to our understanding of visualization design through ``design studies''---projects in which researchers work on real problems with domain experts, including designing a visualization system and reflecting on lessons learned \cite{sedlmair_design_2012}. This form of research is in line with the \textit{research through design} tradition in HCI \cite{zimmerman_research_2007}, where the design of artifacts is accompanied by an intention to produce new knowledge. The design study methodology has led to the creation of useful software tools and valuable lessons reported in the literature through various types of reflections \cite{meyer_reflection_2018}. The typical goal of design studies is not to engage with the practitioner community, however, and the design work involved in these studies is usually conducted \textit{by and for researchers}. Sedlmair et al. \cite{sedlmair_design_2012} noted this clearly, saying ``Our process focuses on design studies as conducted by visualization researchers. Interesting questions rising from this focus include: how does the process generalize to practitioners?'' Based on previous work in multiple design contexts and disciplines \cite{zimmerman_research_2007,goodman_understanding_2011, gray_reprioritizing_2014,stolterman_design_2012,schon_reflective_1983,colusso_translational_2017, cross_designerly_1982,velt_translations_2020} it is unlikely that a process meant for researchers would be sufficiently useful for practitioners. While more recent work has explored important questions about rigor in design studies, including issues of epistemology and broad philosophical approaches \cite{meyer_criteria_2020}, the primary focus is still on knowledge production within and for the research community. 

\begin{figure}
  \centering
  \includegraphics[width=0.7\columnwidth]{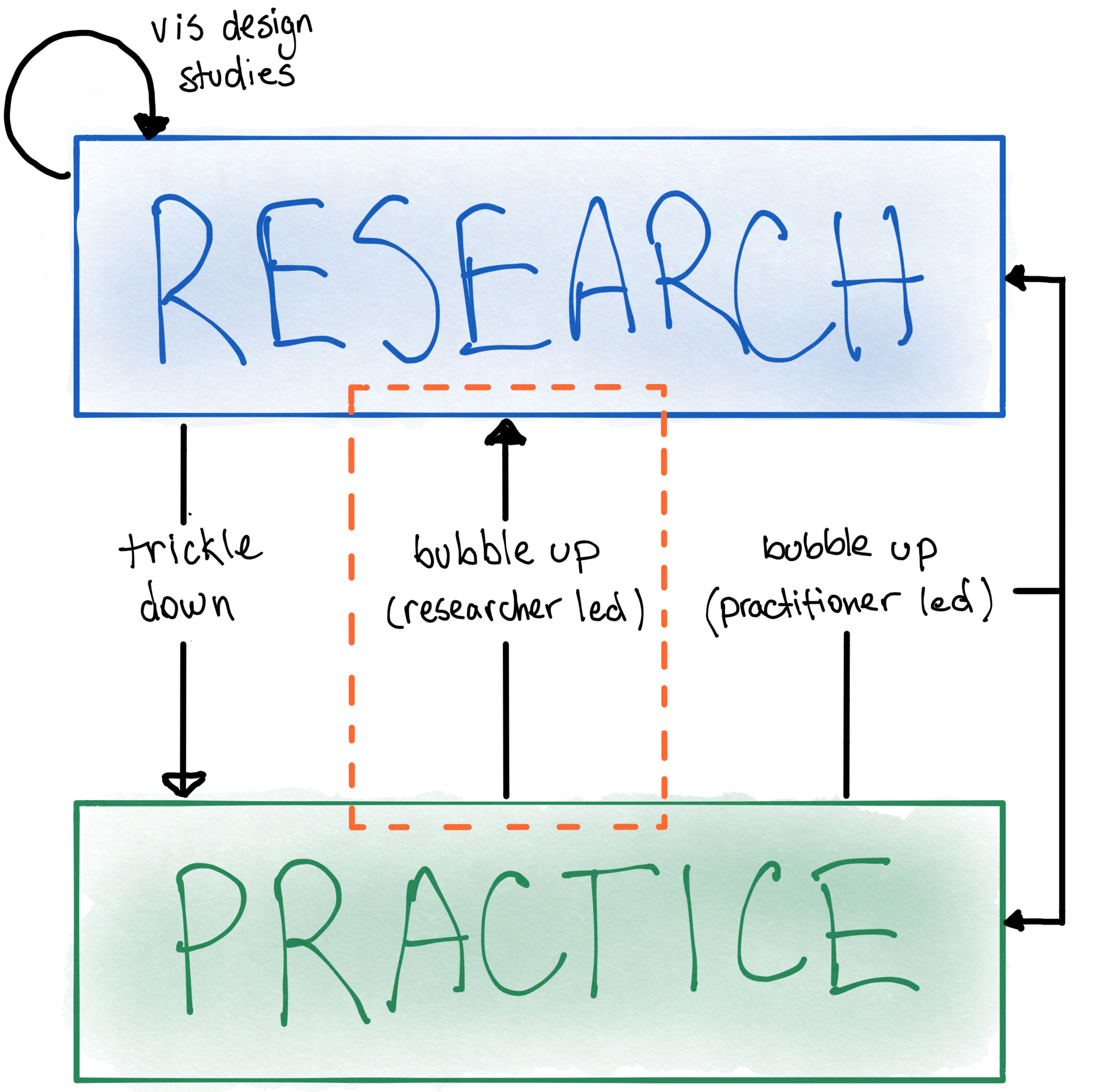}
  \caption{Research-practice relationships. Adapted from Gray et al. \cite{gray_reprioritizing_2014} with the addition of a loop for visualization design studies.}~\label{fig:research-practice}
\end{figure}

HCI researchers have studied and proposed models of relationships among the academic research and professional practice communities (e.g., \cite{gray_reprioritizing_2014,colusso_translational_2019,zimmerman_research_2007,velt_translations_2020}), but the VIS community has not seen equivalent efforts. Figure \ref{fig:research-practice}, which has been adapted from the work of Gray et al. \cite{gray_reprioritizing_2014} engaging with interaction designers, depicts ``bubble-up'' and ``trickle-down'' effects, with the addition of a loop for design studies. The bubble-up effect describes efforts to characterize situated knowledge from practice in abstract, general terms. The trickle-down effect refers to the more traditional view of knowledge transfer, where practitioners opportunistically make use of methods and tools that come from research. Design studies involve researchers doing design work to create a real visualization system, with the primary knowledge output being aimed at the research community. Findings from design studies may trickle-down to practice, although without explicit translational efforts, the transfer of knowledge may not happen. Even with such efforts, there needs to be an accurate understanding of the nature of professional practice to be useful; otherwise, practitioners may view findings as too abstract or not relevant to their work. Knowledge production that bubbles-up from practice to the research community may be led by researchers or by practitioners. The approach taken here is a researcher-led study of professional practice, with the aim of contributing knowledge primarily to the research community. This focus is indicated by the orange box in Figure \ref{fig:research-practice}.

\section{Design Process and Methods}
\label{sec:processmethods}
Many attempts have been made at modeling design processes and methods. Early work, which was promoted most famously by Jones \cite{jones_design_1970} and Alexander \cite{alexander_notes_1964}, proposed design methods as normative schemes that specified activities for designers to perform, including the order in which they should be performed. This early work is often referred to as the ``design methods movement'' \cite{cross_science_1993}. Much of the spirit of the movement was to make design more ``scientific'', aiming for design processes to be rational, formalizable, and generalizable. The goal of this movement ended fairly quickly in failure, however, with both Jones and Alexander rejecting the attempts to provide normative models of design. Gedenryd \cite{gedenryd_how_1998} has noted that ``the failure of these methods is a very solid and widely recognized fact, as is the thoroughness of this failure.'' Scholars have since articulated the reasons for this failure, chief among them being that science and design are fundamentally different activities with different epistemologies \cite{Cross1981}. 

Design researchers moved beyond this attempt to model design methods in normative ways that closely paralleled scientific methods. Löwgren \cite{lowgren_applying_1995} has previously characterized the design methods movement and subsequent work by characterizing three generations of design research. The first generation corresponds to the design methods movement described above. The second generation embraced more complexity in the design process, recognizing the ill-structured nature of design problems. In this generation, the designer's role shifted away from an expert following a scheme to being a ``liberator of the users' needs and requirements.'' The third generation expanded further to recognize the ``specific competence of the designer'', including viewing design as a distinct mode of knowing and thinking. Löwgren argued at the time that the first and second generations had been integrated into CHI, whereas the third was not in the mainstream. 

Although only speculation, as no formal analysis has been done, the evolution of the VIS community may be somewhat similar in the following way: the first generation parallels early work on architectural models and the infovis design space \cite{Card1997}; the second generation parallels the more recent popularity of design studies \cite{sedlmair_design_2012}; and the third generation, in which design is seen as an activity distinct from research and application, with its own forms of knowing and thinking, is not yet as popular---although some scholars have approached visualization design this way (e.g., with respect to design as externalization \cite{davis_design_2019}, the role of aesthetics \cite{moere_role_2011,quispel_aesthetics_2018} and criticism \cite{kosara_visualization_2007}). If researchers are to really understand and influence design practice, we need to expand our scope and philosophy to include characteristics of this third phase.

In the following sections, I describe two studies that were conducted to investigate professional data visualization designers' (1) familiarity with and use of design methods; and (2) design process and knowledge.  

\section{Practitioner Survey}
Prior to the interview study a survey was conducted in which data visualization practitioners reported their familiarity and frequency of use of popular design methods and principles. The development of the survey has been reported elsewhere \cite{parsons_what_2020} so only a high-level summary is given here. The survey took approximately 20 minutes and was fully completed by 87 participants.

\begin{figure}
  \centering
  \includegraphics[width=0.9\columnwidth]{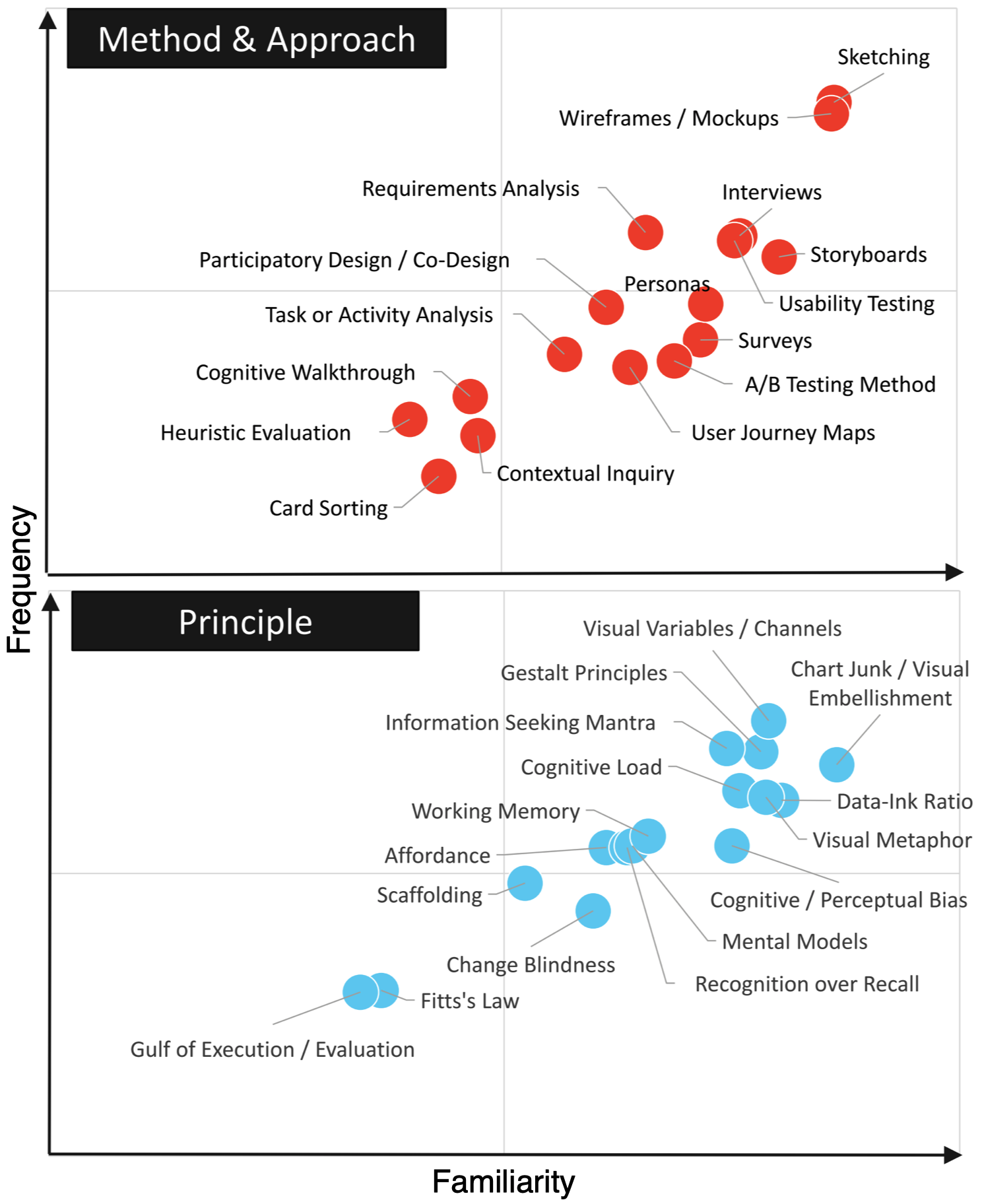}
  \caption{Results of the survey depicting familiarity vs. frequency of use for design methods and principles. Figure reproduced from \cite{parsons_what_2020}.}~\label{fig:2plots}
\end{figure}

Figure \ref{fig:2plots} shows the results where familiarity and frequency are plotted. If an item is in the upper-right quadrant, it is both well-known and frequently used; if an item is located in the lower-left quadrant, it is not well-known and not frequently much. Of course, the upper-left quadrant is not likely to be occupied, yet it still forms a logical possibility. Below is a summary of each category. For the sake of space, only the top two options of the scale (moderately/extremely familiar or often/always used) are reported if 50\% or more of participants selected them. The least familiar and least used items are also identified. The survey data can be found at \url{https://osf.io/9ubrd}.

\textbf{Methods and approaches.} At least 50\% of respondents report being moderately or extremely \textit{familiar} with the following: requirements analysis, interviews, surveys, usability testing, task or activity analysis, participatory design/co-design, sketching, personas, A/B testing, wireframes/mockups, storyboards, and user journey maps. The most unfamiliar methods are heuristic evaluation, with 28 respondents reporting no familiarity. The second and third are cognitive walkthrough and card sorting, with 22 and 21 reporting no familiarity, respectively. At least 50\% of respondents report often or always \textit{using} the following: requirements analysis, sketching, wireframes/mockups, and storyboards. At least 50\% of respondents report never using the following: heuristic evaluation and card sorting.

\textbf{Principles and concepts.} At least 50\% of respondents report being moderately or extremely \textit{familiar} with the following: cognitive/perceptual bias, visual metaphor, cognitive load, change blindness, visual variables/channels, data-ink ratio, chartjunk/visual embellishment, working memory, mental models, information seeking mantra, recognition over recall, Gestalt principles, and affordance. The most unfamiliar principle is Fitts' Law, with 39 respondents reporting no familiarity. At least 50\% of respondents report \textit{using} the following often or always: visual metaphor, cognitive load, visual variables/channels, data-ink ratio, chart junk/visual embellishment, working memory, information seeking mantra, recognition over recall, Gestalt principles, and affordance. At least 50\% of respondents report never using Fitts' Law and the gulfs of execution and evaluation.

\subsection{Discussion}
Many methods and principles were reported as being highly familiar and frequently used. Sketching and wireframing / mockups were significantly more familiar and used than other methods. This is not surprising, as such methods have been shown to lie at the heart of most design work \cite{lawson_how_2006}. The most familiar principle, on average, was chart junk. This result is interesting, although not entirely surprising, as chart junk has been a controversial topic, and considerable debate has taken place in both practitioner and academic spaces (e.g., \cite{borkin_beyond_2016,bateman_useful_2010,borgo_empirical_2012,parsons_data_2020,few_chartjunk_2011}). Other highly familiar and frequently used principles, including visual variables, Gestalt principles, and data-ink ratio, have all been popularized in practitioner-oriented resources for many years.

Multiple methods and concepts were reported as being familiar yet not often used. This includes popular user-centered design methods such as A/B testing, user journey maps, surveys, co-design, and personas. The reason why they are not frequently used is currently unknown. It is possible that some methods are too time- or resource-intensive to use frequently, such as A/B testing and surveys. The familiarity yet limited use of user journey maps may indicate that visualization practitioners are not thinking holistically about a user's journey with a visualization in the way that UX designers often do. The familiarity yet infrequent use of change blindness may suggest that practitioners do not know how to apply such abstract knowledge in their design work---despite a large body of literature on the relevance of change blindness to visualization. Interpreted with reference to the dual gap model shown in Figure \ref{fig:gap}, this may indicate a gap between theory and design instances.

Arguably all of the methods and principles that were reported as low familiarity and low frequency of use are very common in the HCI literature and are well known in the interaction design community. For instance, card sorting, heuristic evaluation, cognitive walkthroughs, contextual inquiry, Fitts' Law, and Norman's gulfs of execution and evaluation have been very well established in the user-centered design literature, and they commonly appear in practitioner-oriented books and other resources. In fact, multiple surveys indicate that heuristic evaluation is consistently one of the most popular and commonly used methods among user-centered designers \cite{vredenburg_survey_2002}. This result suggests that the survey respondents may not have been trained in or familiar with human-centered design practice, despite the fact that heuristic evaluation has been discussed in the visualization literature  \cite{carpendale_evaluating_2008,zuk_heuristics_2006}. Here we can imagine another type of gap, between different communities of research and practice---e.g., between interaction designers and visualization designers. Aspects of this gap have been recently discussed by Walny et al. \cite{walny_data_2020}.

\section{Interview Study}
The survey results provide insight into the methods and principles that practitioners know and use. This is a useful starting point for understanding design practice, but it does not provide an indication of how and why methods and principles are used, nor does it help us understand the design process that visualization practitioners follow. Semi-structured interviews were conducted with practitioners to allow for detailed exploration and probing of their design process and use of methods. The guiding research questions for the interview study were: (1) How do practitioners engage in their design processes and what strategies do they use? and (2) How are various forms of design knowledge used in practitioners' design practice?

Regarding the first question about design process, multiple sub-questions were used to guide the discussion: (a) how do practitioners start their design process?; (b) when working through a design process, how do practitioners assess their progress, including how to determine if they are on the right track and what steps to take next?; (c) what is the structure of the process---e.g., is it systematic, logical, step-wise, iterative, linear?; and (d) how do practitioners make decisions regarding which chart types and visual encodings to use within their process? There were also sub-questions regarding the forms of design knowledge: (a) do practitioners rely on any high-level theory to guide their design work? (b) how are methods and principles used in practitioners' design work? (c) are there methods and principles that are more useful than others? and (d) are practitioners aware of the history and development of popular visualization design concepts? To address these last questions, participants were asked about the two of the most popular and widely used principles according to the survey---chart junk and visual channels (see Figure \ref{fig:2plots}). If participants didn't recognize the concept initially, a definition was provided along with examples to ensure a shared understanding before moving forward. After doing this, every participant stated that they were familiar with these two concepts. 

\subsection{Method}
20 data visualization design practitioners participated in the interviews. Recruiting was done via social media, the DataVis Society’s Slack workspace, the InfoVis email list, and personal networks. To mitigate sampling bias, practicing professionals and agencies were sought out through web searches and more than 200 individuals and more than 30 agencies were contacted to participate in this study. 18 of the 20 interviewees completed the survey prior to the interview. The other 2 responded to the request for interviews but did not fill out the survey.

Interviews were conducted remotely via videoconferencing and were transcribed. Interviews took place during July and August of 2019. The protocol for this work consisted of two primary topics: (1) design process and strategies, and (2) design methods and principles. Each topic had an open-ended lead question, intended to establish rapport and make participants comfortable talking about their experiences, and a backup question that was more targeted in case participants were not sure how to answer the initial one. Each topic also had multiple follow-up questions that were asked roughly in the same order but depended on the previous answers that were provided. Both the survey and interview were approved by the IRB at Purdue University.


\subsubsection{Participants}
Participants came from multiple countries, although mostly from the US and Northern Europe (individual countries are not listed to maintain anonymity). Participants had a range of job titles, although all self-identified as data visualization designers. Aside from P7 and P5, who reported spending about 1/3 of their time on data visualization and the rest on other data or UX/UI issues, all participants reported spending the majority of their time on data visualization related work. Participants also worked in a range of practice contexts, including freelance work, governmental work, and work across both small and large companies. Table \ref{tab:participants} lists the self-reported details of our participants, including their job title, location, job context, experience in years, highest degree attained, and gender.

\subsubsection{Analysis}
All interviews lasted between 50 and 90 minutes, with the average at 67 minutes. Collectively, the interviews were approximately 23 hours. All interviewers were conducted by the author; occasionally a student sat in to observe. Each interview was automatically transcribed using a speech-to-text transcription service called Temi. Transcripts were then examined and errors in the automated transcription were fixed.

The transcripts were inductively coded by the author, following standard processes for thematic analysis \cite{braun_using_2006}. The coding process took place over a period of one year, with sustained efforts at a few points in time. Coding was done fully by the author, but five other researchers with whom the author collaborates regularly (including one professor, one PhD student, two MS students, and one undergraduate student), agreed to check the codes at different points in time. Each person independently examined approximately 10-15 minutes of 10 transcripts, and meetings were held to check the codes and assess their accuracy. A software platform called Dovetail was used to facilitate the coding. 

This thematic analysis occurred in three main stages. First, while getting familiar with the transcripts, using active, repeated reading \cite{braun_using_2006}, the author took notes of salient portions of the transcripts. Second, the author relied on the notes and subsequent readings of the transcripts to identify initial codes relating to the questions in the original interview protocol (see above). This open coding phase, which required multiple passes through the transcripts, resulted in approximately 60 codes being generated. The codes were then examined in context and compared to one another, leading to some being removed or re-labeled and others being merged. This stage was repeated multiple times as the codes were iteratively defined. Third, the author closely examined the codes, attempting to cluster them and generate themes and sub-themes. This stage also occurred repeatedly, as the themes were refined with the criteria of internal homogeneity and external heterogeneity in mind \cite{patton_qualitative_2014}. Finally, the author defined and named the themes, attempting to capture the essence of each.

\begin{table*}
\centering
\begin{tabular}{ |c|c|c|c|c|c|c| }
\textbf{ID} & \textbf{Job Title} & \textbf{Location} & \textbf{Job Context} & \textbf{Exp. (yrs)} & \textbf{Highest Degree} & \textbf{G} \\ 
\hline
P1 & DataVis Journalist/Designer & Europe & Freelance & 5-7 & M & M \\
P2 & Sr. UX/DataVis Designer & N. America & Government & 5-7 & M & F \\
P3 & Sr. DataVis Dev & N. America & Large company & $>$10 & D & F \\
P4 & Data Communicator & N. America & Government & 2-4 & M & M \\
P5 & Data Storyteller & N. America & Small company & $>$10 & M & M \\
P6 & DataVis Engineer & N. America & Small company & 8-10 & D & M \\
P7 & DataVis/UX Designer & Europe & Freelance & $>$10 & M & M \\
P8 & Sr. DataVis Designer & N. America & Large company & $>$10 & D & F \\
P9 & Sr. UX Design Lead & N. America & Large company & 8-10 & D & M \\
P10 & DataVis Designer/Dev & Europe & Freelance & $>$10 & M & M \\
P11 & DataVis Designer & Europe & Small company & 5-7 & M & F \\
P12 & DataVis/UX Designer & N. America & Large company & 2-4 & M & F \\
P13 & Graphics Editor & N. America & Journalism & 8-10 & B & F \\
P14 & DataVis Designer & N. America & Freelance & 5-7 & B & F \\
P15 & DataVis Designer & N. America & Large company & 8-10 & M & M \\
P16 & DataVis Designer & N. America & Small company & $>$10 & M & M \\
P17 & DataVis Designer & Europe & Freelance & 8-10 & B & M \\
P18 & DataVis Designer & N. America & Freelance & 5-7 & B & M \\
P19 & DataVis Designer & Europe & Freelance & 5-7 & M & F \\
P20 & Data Architect & N. America & Small company & $>$10 & M & M \\
\hline
\end{tabular}
    \caption{The 20 participants and their self-reported characteristics: job title, location, job context, years of experience, highest degree (Bachelor's-B, Master's-M and Doctoral-D) and gender. ~\label{tab:participants}}
\end{table*}

\subsection{Findings}
Findings are presented here with respect to the two guiding research questions---the first focusing on design process and strategies, and the second focusing on design knowledge.

\subsubsection{Design Process}
Participants described a wide variety of processes, with no discernable consensus on specific steps or procedures that were taken. Participants described doing many of the same things, including data cleaning and wrangling, brainstorming, talking to clients, sketching and prototyping, user testing, and others. These activities were commonly performed, yet took place with little regularity in terms of where they occurred in the process. Some participants started their process by trying to understand the data; others by trying to understand the users' needs or the clients' goals; some would wait to visually represent the data, while others would do so right away as a means of problem framing. The themes that led to this assessment are discussed below.


\textbf{Data, Users, or Clients First?} 
When asked about how they begin their design processes, many participants referenced a data-first kind of mentality. For instance, P12 stated ``\textit{I feel like I can't really start designing until I have some kind of data that I can play with, because my design is going to be very dependent on the shape of the data.}'' In a similar fashion, P1 noted ``\textit{I don't really think that there's a typical design process or typical data visualization project. But usually the first thing is get the data.}''

Other participants talked about more of a people-first approach, with P7 saying ``\textit{Typically, I think it would start with some kind of a product owner, or a CEO or somebody, having some kind of rough understanding already or thinking what it is that they want to work on or accomplish. Then at that point I try to somehow dig out what the actual end-user requirements are.}'' P9 described initially focusing on the client goals: ``\textit{I guess I would spend some time with whoever is commissioning it [\ldots] to really understand what is the use case, what questions are they looking to answer? And ultimately, I always start with three questions---all context. So who's it for, what's the purpose, and how's it going to be delivered?}'' P18 described a similar mentality, stating ``\textit{So, high level, I'd like to understand who is pushing for this initiative. If it's a dashboard, who are my champions? Get those stakeholders together and try to do some research to understand how they see the world a little bit.}'' P5 described initially prioritizing people over data, saying they ``\textit{begin with trying to understand the audience and the action that my clients are wanting to reach. So I don't begin necessarily with the data.}'' 




\textbf{Assessing Progress.} 
Participants largely did not describe following specific steps or procedures to help them move through their process. Rather, participants often talked about their intuition or experience guiding them. Sometimes this was implicit, such as in this example with P5: ``\textit{How do I know I'm going in the right direction? You know, I think I look for the quick reaction I can get from what the visualization is showing me. When I look at the data I might say, wow, that's a really compelling trend. Or, those disparities really stand out to me. But sometimes when you look at it as a visualization, you know what, it just doesn't show that way. For whatever reason.}'' Participants also explicitly talked about some form of ``intuition'' or ``gut feeling'' that guided them through their process, with P6 stating ``\textit{I think part of it [assessing progress] is intuitive. You know, I have a sense for often what is the piece of the thing I'm trying to communicate or the question that I'm trying to answer with the data. [\ldots] it's like you keep trying until you finally get the piece that answers that question.}'' P19 described the role of experience as ``\textit{it's a feeling of having this sort of mental model of the data, and it's something that I feel builds up through experience, having used these kinds of visual forms before. And it's like a feeling that I feel like, I think I can make this visual form look most interesting.}'' In response to how they assess their progress and determine how to move forward, P13 stated ``\textit{I don't know that. It's kinda like this gut feeling. And I know that's not very scientific, but I can look at a chart and be like this just isn't showing what I want it to show---it's kinda like a gut check.}'' P11 simply stated ``\textit{I feel like mostly this is intuitive for me right now.}''



Aside from their own intuition, participants also talked about evaluating their work by involving other people to help determine how to move forward. This typically involved showing their work to either clients or colleagues and friends. P6 describes their view on how their own judgment often works well, but they also like to test with users: ``\textit{I have become quite a good judge of `is this thing going to convey the message that I think it will?' But of course you know, the ultimate test of that is to show it to someone and say, `what do you see in this?}''' P16 described something similar saying ``\textit{there's a lot of discussions with the client showing them things and getting their feedback [\ldots] I also will often check with you know, friends, colleagues, my wife, whatever, to say does this communicate what I wanted to communicate? What do you think it says?}''


\textbf{Planning in Action.} In addition to not relying on predetermined design processes, many participants described a kind of situated planning that often relied on forms of experimentation and responding to what was happening in the moment. For instance, some participants talked about creating visual representations almost immediately in their process, either via sketching or throwing the data into software, before trying to figure anything else out, often just to get a ``feel'' for things. P14 exemplifies this kind of planning-in-action ``\textit{so when I get the data, the first thing I do is I take a look at for each dataset, the metadata, the attributes, and then I kind of just start thinking about what might be interesting about that dataset or interesting questions to ask that dataset. And then once I formulate some questions, that's when I start using---recently I've been using a lot of Vega Lite to do some of the charting, just to try and see if I could answer those questions. And I'll keep trying at it until I get a satisfactory answer.}'' Additionally, P17 described forming ideas about visualizations in their mind even before working with the data: ``\textit{once I start talking with a client, or in an initial conversation about a project, usually I already get some ideas in my mind about how I could approach this or how I can visualize it.}'' With this kind of responsive, situated planning, the designer does not have a pre-determined vision of where the process is going, but rather engages with the situation in a conversational manner. P19 exemplifies this when describing her iterative sketching process, saying \textit{``I call it designing with code. At the start, I don't know my color palettes, I just know very roughly, in an abstract sense, how I want to place my data on the screen. Is it going to be radial, is it going to be circles, rectangles, curves, whatnot. And only by iterating a lot with different things, you know, going from circles to rectangles or using size, or maybe variable x or y, do I kind of [\ldots] slowly iterate my way to the final design that I feel I'm okay with.''} P1 also illustrates this way of knowing and acting, stating ``\textit{There are no really real hard steps in my process that I say, well, okay, I'm gonna sit down and look what I've done and then think about what direction I should take. It's more like an organic and intuitive kind of process of gradually getting new insights and new ideas and trying things out and getting rid of things.}''


\subsubsection{Design Knowledge}


Participants were asked about their use of various forms of knowledge, including theories, principles, methods, and guidelines. None of the participants reported relying on any high-level general theories such as distributed cognition, activity theory, dual process theory, information theory, and the like. P6, who has a PhD in information visualization and now works as a practitioner, mentioned some academic frameworks like the Design Activity Framework, the Nested Model, and Design Study Methodology. When asked specifically about how they influence his design work, P6 noted that they've helped him think ``\textit{very explicitly about what are the tasks, why is someone sitting down to use my visualization? What's the context that they have? And then what are the decisions that I make as a designer to try to work within that context.}'' When asked more specifically about any kinds of frameworks or taxonomies they use that might help structure the design process, P6 stated ``\textit{I think I use Tamara's language [\ldots] I don't remember which taxonomy it is, but like query versus look up versus browse and search versus explore. I use that language a lot, but I often find that it's almost too complicated. And trying to create a taxonomy for tasks---it's hard and it's so abstract [\ldots] they're an interesting way of thinking about design, but I don't really consider them useful in part because there's just so much variability in how people think.}'' This quote exemplifies the challenge of bridging the gap between theory and practice as described by the dual gap model (see Figure \ref{fig:gap}), even for someone with a PhD in the field.

In reference to more intermediate-level knowledge, however, participants referenced many methods, principles, and guidelines that they relied upon. Commonly mentioned concepts included affordance, data-ink ratio, cognitive load, data types, Gestalt principles, chartjunk, visual variables, and mental models. Participants easily recalled and discussed how they relied on these concepts. For instance, P4 discussed cognitive load and mental models this way ``\textit{I'm often thinking about cognitive load because I often think that my job is to simplify and clarify. I'm often thinking about mental models because I often think that within government we think about things in one way. And citizens, residents, users for the general public often think about them in a different way.}'' P1 described the importance of thinking about data types: ``\textit{the data itself will guide you, and it will also limit what you can do. For example, if you don't have a geographical component in your data, you cannot make maps [\ldots] if you have numbers and time dimensions in your data, this will determine what kind of visualizations you can make.}'' Participants sometimes described how they used these concepts together, such as P18 ``\textit{I like to use the gestalt principles of proximity instead of closure, if I can get away with it \ldots and Tufte's data-ink ratio. If I don't have to draw a box around something and I can relate them to one another just by their proximity, I try to leverage that.}''

\textbf{Precedent and Inspiration.} Regarding the use of precedent to inspire, P19 illustrates this by saying ``\textit{With the project in the back of my mind, I start browsing my Pinterest boards for inspiration. Anything that kind of seems to click with what I have in mind, I will then put that into a client sort of mood board. And once I feel like I have enough visual stimuli, I will kind of put that on the screen while I have my iPad in front of me and I'll just start drawing out ideas.}'' Some participants described going beyond data visualization to look for inspiration, such as P14 saying ``\textit{I try to then look for inspiration that's outside of datavis. And so just like, I just really like art. And different forms of art in nature.}'' P13 similarly stated ``\textit{If it's a bigger thing and I want to try to be more ambitious, I'll kind of look around---I have a bunch of books on my desk I use for inspiration, or sometimes I'll go to art museums around here.}'' Especially when faced with difficult situations, or when they felt stuck in their process, participants looked for inspiration rather than models to guide them. P20 exemplifies this when describing struggling with an existing design that was not very good ``\textit{I just kinda went into sort of a hysterical, `inspire me' mode, and spent the better part of two days reading and looking at visualizations and trying to find things that I hadn't seen. And I ended up stumbling on---you know what a waffle chart is? [\ldots] And so I stumbled on one of those, and I hadn't seen it before, and it was visually interesting. And so that, combined with with a better color palette than they had before, turned into a pretty compelling result that they liked and then got me out of the woods.}''

\textbf{Scientific Knowledge.} Most participants discussed, at least implicitly, the important role that scientific knowledge had for them, although not in a prescriptive or normative way. P19 exemplifies this when discussing how perceptual rankings can be useful in supporting their process in the background: ``\textit{I think that was very important for me really to sort of get a ground level of knowledge knowing that people are better at comparing lines than they are at angles [\ldots] it's good that somebody articulates that. It's really sort of the fundamental things you should know to make a good data visualization, but it doesn't teach you any way how to go from one specific dataset to a design---but it kind of gives you a lot of the handholds to keep in mind while you are trying to figure out how to go from data to design.}''

\textbf{Choosing Chart Types.}
Participants were asked how they choose chart types and visual channels in an attempt to surface some of their decision making strategies. Participants rarely described any kind of rational choice approach resembling a cost-benefit analysis or a systematic consideration of tradeoffs. P3, who has a PhD in information visualization, described relying on standards with a focus on readability: ``\textit{Something that I think I always try to consider [\ldots] is chart readability. So going back to the Cleveland and McGill standards. I feel like my role is often just making sure that whatever we're doing, we're kind of meeting that threshold of readability and usability.}'' P3 then noted there are other important factors to consider, such as keeping the client engaged: ``\textit{if you throw a bunch of bar charts on a page with no variation, maybe each of those is very effective in terms of that low level of perception. But if a client gets bored and leaves, then none of them are effective.}'' In discussing rankings of visual channels, P14 noted ``\textit{The very first thing I think of is [\ldots] what is a good visual metaphor to clue people into what I'm trying to say. And so actually that takes precedence over the ranking.}'' P15 described the selection of chart types and encodings as more intuitive: ``\textit{I know from being in the field what's supposed to work well in certain situations and what's not supposed to work well. And you know, I think I chalk some of that up to intuition. It's like some kind of combination of intuition and experience.}'' Others described relying more on precedent for inspiration, with P7 stating their choice involves ``\textit{looking at books or the web for inspiration.}''

Others based chart types and encodings on a communicative intent, as in what message they want to convey. For instance, P19 stated ``\textit{Sometimes it's kind of obvious if I, for example, had a project where I visualized flow, this is money coming in and out of something. So Sankey felt so natural as a base form to start with and then iterate from there. Or when you have a network, when you want to show connections, some sort of network seems like the obvious way to try and start things. But other projects, it's not clear at all. Usually then I try and look back and say, what is the main thing that I want to show?}''

\subsubsection{Context of Professional Practice}
As reported in Table \ref{tab:participants}, participants in this study had diverse contexts relating to their professional practice, including their job roles, geographical location, and organizational context. It was not an original aim of this study to investigate the role of these factors in their practice, so no formal analysis was done. However, some observations can be noted here. Participants all self-identified as designers, with most practicing in fairly broad range of applications and domains. P13 was the only participant operating entirely in a journalistic context, and described the role of editor oversight, noting that design decisions ``\textit{usually have to get approved by an editor}'' and that there are ``\textit{editors in my path to production that will stop me if it's a really bad idea.}'' There may be similarities to manager oversight in company settings, but no such statements were made by participants. Freelancers tend not to have such kinds of oversight---but this does not mean they receive no critical feedback on their work. Most participants described how they seek out feedback from peers, co-workers, or even informally from friends and family members. There was no discernable difference in geographic location, although participants were limited to North America and Europe, and differences may exist in a more diverse sociocultural sample. 

\section{Discussion}
The findings of this work suggest that practitioners rarely use any kind of logical methodology to guide their design process. Rather than following pre-determined plans or process models, participants described a more situated kind of knowing and acting, where planning happened in the moment and preceded action only very locally in time. The majority of participants explicitly referred to ``intuition'' or ``gut feeling''---which are typically synonymous with experience---as guiding their process and helping them assess whether they were going in the right direction. Participants described looking for inspiration from precedent visualizations---or even from art and nature---to help them move through their process. Bigelow et al. \cite{bigelow_reflections_2014} have found that visualization designers do not like tools that enforce a particular process, as the tools can be inflexible and create extra work. Our work seems to confirm this finding, and also extends it in important ways. Our finding that practitioners do not follow prescriptive processes goes beyond issues of flexibility and tool use, pointing to a distinct epistemology of practice at a deeper level. This raises questions regarding the nature of design knowledge for research and practice, including where differences and similarities may lie, and how these distinct ways of knowing and acting can be recognized and leveraged for strengthening knowledge production and use between the two communities.



Design researchers in other fields have discussed these issues, and visualization research may benefit from adopting conceptual and methodological contributions that they have made. For instance, Lawson has described how expert designers largely rely on precedent and gambits (or `tricks') to guide their work \cite{lawson_schemata_2004}. Rather than being guided by models, experts use gambits drawn from patterns of prior experience that have recognizable properties and solutions. This pattern-matching strategy is seen across a wide variety of domains involving expert decision making \cite{klein_sources_2017}. Empirical investigations show that people instead rely on patterns of experience to make in-the-moment judgments that are often perceived of as ``intuition'' \cite{beach_why_1993}. Cognitive scientists have found that people rely on artifacts in the environment, rather than plans in their heads, to serve epistemic functions and aid in decision-making in the moment \cite{kirsh_interactivity_1997,kirsh_distinguishing_1994}. From a sociological perspective, Suchman \cite{suchman_plans_1987} has argued that all meaningful action is situated, depending in essential ways on material and social circumstances that cannot be adequately planned for ahead of time. It is only after the fact that people ascribe rational plans to their activity.

Schön famously described designers as ``having a reflective conversation with the materials of the design situation'' \cite{schon_designing_1992,schon_reflective_1983}, meaning that designers actively construct and re-construct the structure of the situation, determining what to attend to in the moment rather than ahead of time. This conversation unfolds through a blending of active sensory appreciation of the materials and objects in play and the knowledge and experience of the designer. The designer makes judgments and performs certain actions, which are then reflected upon in situ to determine their suitability. In following this process, the designer does not have a pre-determined vision of where the process is going, but rather engages with the materials in a conversational manner. In our interviews, participants described this kind of `conversation' they would have with respect to identifying appropriate visual encodings, chart types, and even framing the problem space through active experimentation with the design materials. More investigation into how designers converse with their materials could be a valuable line of research for the visualization community.

The kind of reflective, situated knowing and acting that participants described is not captured well with existing visualization design models. For instance, process models (e.g., \cite{sedlmair_design_2012,mckenna_design_2014}) often suggest a particular sequence of steps that should be followed. Even if they allow for iteration, or moving between different stages, they still promote a sequence and do not capture the ``conversation'' that practitioners have. The systematic approach to design (see Section \ref{sec:processmethods}) may be good for structuring research investigations, but it does not appear relevant for professional practice. There is a need for visualization design research to capture this kind of knowing in action that practitioners engage in as part of our understanding of how data visualization practice is done.

In reference to the kinds of knowledge practitioners rely on, they appear to have little use for high-level theories, particularly if they had a prescriptive focus. Although abstract knowledge structures like task taxonomies are popular in the visualization literature, participants did not report using them. Even P6, who has a PhD in information visualization, describes task taxonomies as ``abstract'', saying ``\textit{I don't really consider them useful}'' in design practice. P3, who also has a PhD in visualization, described such frameworks as ``\textit{not something that I would actively use.}'' What was found, however, was that practitioners rely often on individual concepts and principles that could be called to mind in the moment of a design situation. Participants could easily recall important concepts like visual channels, data types, cognitive load, chartjunk, affordance, and others. Participants described using these concepts `often' or `all the time'. Although participants rely on these individual concepts regularly, they did not do so in a prescriptive way. Rather, they would call them to mind in the context of the `conversation' that was taking place.

These findings align with research done in other design disciplines. Based on studies done in various contexts, Stolterman \cite{Stolterman2008} suggests that design practitioners are inclined to appreciate and use four forms of knowledge in their work: (1) precise and simple tools or techniques (e.g., sketching, prototypes); (2) frameworks that do not prescribe but that support reflection and decision-making; (3) individual concepts that are intriguing and open for interpretation and reflection on how they can be used; and (4) high-level theoretical and/or philosophical ideas and approaches that expand design thinking but do not prescribe design action (e.g., human-centered design). The findings of this study appear to be in line with this suggestion. Most participants described using simple techniques like sketching and prototyping as core aspects of their process. Not many participants mentioned frameworks, but P3 and P6 described frameworks as influencing their thinking, although not in a prescriptive manner. Individual concepts were very commonly described as important in supporting our participants' design work. Participants did not often describe high-level philosophical ideas explicitly, but influences of human-centered design were often apparent when they described their process.

It was clear that the ways in which practitioners make decisions during their process relies on various types of \textit{design judgments} rather than logical decision making strategies. Although rational decision-making processes are often viewed as ideal for dealing with complex situations, research shows that expert designers rely on situated forms of judgment. However, a detailed analysis of design judgment is outside the scope of this paper and has been presented elsewhere \cite{parsons_design_2020}. 



\subsection{Modeling the Design Process}
Based on the findings discussed above, it does not seem possible to usefully model the design process of practitioners. Participants described many kinds of dependencies---e.g., wanting to explore the data before selecting chart types, or creating a prototype before doing testing---but these dependencies were not at the level of steps or stages that were shared across participants. However, there seem to be common features of characteristics of practitioners' design processes that can be articulated. For instance, most participants described continually drawing from various sources of inspiration, precedent, and methods and principles in ways that reflect knowing-in-action and Schön's notion of having a reflective conversation. 

The inability to develop a model of practitioners' process is not unique to data visualization. Design researchers have previously noted how design processes can be too complex and varied to model in a generic way \cite{lowgren_thoughtful_2004,nelson_design_2012}. Löwgren and Stolterman \cite{lowgren_thoughtful_2004} have argued that there are fundamental aspects to the design process that cannot be separated from it---they take place continually and not in any particular stage of the process. Examples of these recurring aspects include jumping from the big picture to specific details, and dealing with various design dilemmas and tensions. Participants in this study described engaging in these activities at various points in their process. Participants also described doing at least 3 things in parallel throughout their process: (1) framing and structuring the problem or opportunity space; (2) generating ideas and prototyping; and (3) testing and evaluating. While participants described placing more emphasis on these at different points in the process, it was not to the extent that they could be modeled generically. These 3 things appear to happen in parallel and feed one another at multiple points along the way. These characteristics appear more similar to a design sprint or agile model than a structured linear model.

First, regarding framing and structuring, a fundamental element of all design is shaping the problem or opportunity, including identifying which things to attend to and which to ignore. Some participants described clients coming to them or their company and wanting something---yet not being entirely sure what they want. To engage in this kind of framing and structuring, participants described engaging in activities such as requirements gathering, exploring and making sense of the data, talking to users and clients to understand their needs, and trying to figure out what message or story should be communicated from the data. Although problem framing tends to dominate the early part of the process, previous work in other design disciplines shows that is is rarely done only at the beginning \cite{schon_designing_1988,Goel1992}. 

The second thing all practitioners described doing is generating ideas and prototyping. This kind of activity tends not to be the very first or last thing done in the process, although it can happen very close to the beginning or end. Some practitioners described sketching or doing low-fidelity prototyping work very early on, as a generative exercise before having a clear problem frame in place. Most participants described doing prototyping in an iterative fashion, where prototypes served various functions, including helping to frame the problem, filter design ideas along different dimensions, and communicate design ideas for feedback \cite{lim_anatomy_2008}. Participants described multiple techniques for generating and refining ideas, including sketching and wireframing, looking for inspiration in other visualization work or elsewhere, drawing on principles regarding visual communication, cognitive load, or perceptual tasks and visual encodings, and others. Prototyping and idea generation were described as taking place throughout the process, often right up until the visualization tool was shipped to the client. 

The third thing all participants described doing is testing and evaluating their work. Although evaluation is not as common early in the process as it is later, participants sometimes described getting quick feedback on sketches or other low-fidelity prototypes early in the process, often as a means of helping with framing and determining how to move forward in the ideation process. 

The way that participants described their design process involves these three activities occurring in parallel, in ways that were often informing one another. Furthermore, participants described drawing from many different methods and principles as they moved through their process---as means of structuring the problem space, generating and refining ideas, assessing progress, and evaluating design ideas---in ways that were opportunistic and involving just-in-time learning and application. Future work may be able to develop models---other than process models---that capture these and other characteristics of how practitioners design data visualizations. 


\subsection{Revisiting Research-Practice Relationships}
Findings from the interviews can be discussed and interpreted in terms of the two models presented earlier---the dual gap model \cite{velt_translations_2020} and the bubble-up/trickle-down model \cite{gray_reprioritizing_2014}. The gap between high-level theory and specific design instances unsurprisingly exists in the practitioner community. Most practitioners expressed an appreciation for academic research but did not express much interest in high-level theories. They did express interest in intermediate-level knowledge, especially principles of perception and visual encoding, visual and graphic design principles, and guidelines like the information seeking mantra and the data-ink ratio. 

The gap between academic research and professional practice can be examined with more nuance using the bubble-up/trickle-down model shown in Figure \ref{fig:research-practice}. In the interviews there was some evidence of trickle-down effects taking place, where findings from academic research were being used in practice. Examples include work on perceptual tasks and visual channels from researchers like Cleveland and McGill \cite{Cleveland1985} and Munzner \cite{Munzner2014}. These efforts may be amplified by recent efforts like the Multiple Views Blog \cite{noauthor_multiple_2019} and ongoing conferences and workshops.

There was evidence of knowledge being spread among practitioners in a way that details of practice `bubble-up' to something more general. Many practitioners have created books and blogs to offer practical design tips and guidelines for other practitioners. Names of some well known practitioners were mentioned multiple times by participants as serving in this role (e.g,. Manuel Lima, Nadieh Bremer, Stephanie Evergreen, Alberto Cairo, Giorgia Lupi, Moritz Stefaner, and others). Participants did not describe efforts to bubble-up knowledge to the research community, although there are some popular works by practitioners which have been discussed by academics previously (e.g., \cite{bihanic_new_2015,lupi_dear_2016,bremer_data_2021,kirk_data_2012,Cairo2016}). This may change with the recent creation of conferences and workshops, although special effort will likely be required to do this in ways that both practitioners and researchers benefit from.

\begin{figure}[h]
  \centering
  \includegraphics[width=0.7\columnwidth]{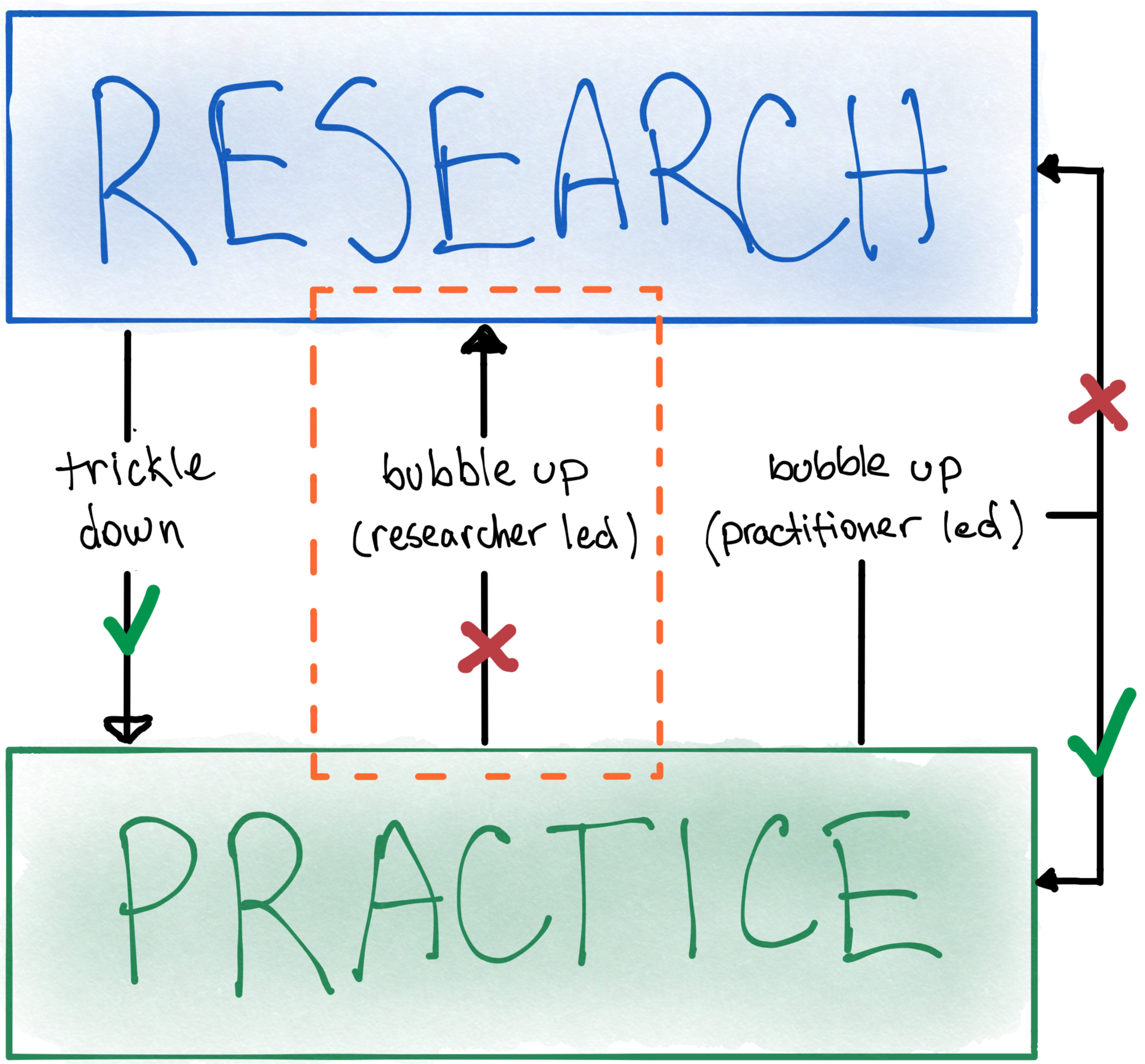}
  \caption{Revisiting the trickle-down/bubble-up model adapted from Gray et al. \cite{gray_reprioritizing_2014}. Check marks indicate where there is at least some evidence of knowledge transfer and x marks indicate there is not very much.}~\label{fig:modelrevisited}
\end{figure}

This current work took a researcher-led approach to studying practice with the aim of bubbling-up knowledge for the research community. This kind of researcher-led inquiry, where design practice is investigated on its own terms, currently makes up a minority of the literature on visualization design. This work tentatively maps some of the theoretical and conceptual landscape to facilitate future inquiry of this kind. Figure \ref{fig:modelrevisited} depicts where evidence of knowledge transfer is happening between the research and practice communities. Although it somewhat of a simplification, it may indicate where future research can focus.

\section{Limitations and Future Work}
As with any study, there are limitations of this work that must be kept in mind when interpreting the findings. Regarding the sampling strategy for recruiting participants, there may be bias due to reliance on convenience sampling---it is known that participants who volunteer are likely to be more invested in the topic and may feel strongly about certain outcomes \cite{moore_statistics_2006}. There may also be a lack of representativeness with respect to sociocultural factors, as all participants were from North America or Europe. Because participants self-identified as designers, other identities (e.g., engineer, developer) may not have been represented to an equal extent. Finally, as there is no clear picture of the population of data visualization practitioners, the sample's overall representativeness is unknown, and findings should be generalized with appropriate caution. Future studies could focus on contextual aspects of design practice that were not addressed here, including levels of experience (e.g. novice, expert), background and training (e.g., computer science, design, art), organizational context (e.g. design studio, large corporation) or work context (e.g. large design team, sole designer).  

Aside from the sample, there are limitations with respect to the methods used. The study is based on self reported recollections from participants, which may be subject to known biases and errors. Future research should use multiple methods for investigating design practice, including controlled experiments and ethnographic field work. Longitudinal studies could help understand variations in design processes and their underlying influences.

The findings of this work support an emerging interest in the literature on understanding data visualization practice  \cite{bigelow_reflections_2014,bigelow_iterating_2017,mendez_bottom-up_2017,hoffswell_techniques_2020,walny_data_2020,parsons_what_2020,alspaugh_futzing_2019}. Future work can investigate modes of knowledge sharing between the research and practitioner communities, including ways to promote and strengthen bubble-up and trickle-down effects. Despite the existence of design models in the visualization literature \cite{munzner_nested_2009,mckenna_design_2014,sedlmair_design_2012}, it is not clear how accurately they capture the design process that researchers follow, and how similar it may be to the process of practitioners. There may be value in engaging with the research through design literature (e.g., \cite{zimmerman_research_2007}) to strengthen the theoretical landscape underlying research, design, and practice. Additionally, investigation into the ways in which academics learn from and make use of practitioner resources would make an important contribution. 




An implication of this work for data visualization education is the attitude educators take towards preparing future practitioners to handle the complexity of real-world practice. Specifically, this work and previous work in other design contexts shows that training should focus on \textit{preparing designers for action} rather than attempting to \textit{guide designers in action} \cite{Stolterman2008}. This may seem like a subtle point, but the implications are significant. Rather than teaching students to follow processes to achieve good design outcomes, effective design education should focus on the competencies needed to operate skillfully in context. These include knowing how to frame problems, draw from precedent, recognize patterns to make informed judgments, among other skills that can be relied on in the moment. If students are not prepared to face the complexity of real-world practice, prescriptive models and guidelines will not effectively guide them through such situations. However, adopting this attitude requires a recognition of the epistemology of professional practice as being distinct from the epistemology of academic research \cite{schon_designing_1992}. Future work can investigate ways of training data visualization designers in relation to studio practices \cite{vorvoreanu_advancing_2017,ridley_evaluating_2020}, such as the use of critique \cite{Brath2016,fox_surfacing_2020}, and the development of a design identity \cite{gray_building_2020}. 

\section{Conclusion}
In this paper, I have argued that more practice-led research is needed if the visualization research community wishes to understand and influence professional design practice. Findings from a survey and interview study were presented, summarizing how practitioners describe their design process and the kinds of knowledge they value and use. Findings suggest that practitioners do not follow prescriptive processes, and instead rely on precedent, experience, and intermediate-level knowledge to guide them in a situated conversation with the design situation. Findings suggest that strengthening relationships between the research and practice communities requires an understanding of the epistemology of practice and how it differs from that of research.

\acknowledgments{
The author would like to thank students from the DVC Lab, including Ya-Hsin Hung, Ali Baigelenov, Erica Chadwell, Connor Schrank, Ian Carr, and Labonno Zaman. The author would also like to thank Colin M. Gray for stimulating discussions about design theory, and the anonymous reviewers from two venues for providing helpful suggestions on earlier drafts. Many thanks are due to the participants in this study, who devoted their time to make this research possible. This work was supported in part by a grant from NSF (\#1755957).}

\bibliographystyle{abbrv-doi}

\bibliography{template}
\end{document}